# Quantum secret key encryption algorithm based on quantum discrete logarithm problem

Chien-Yuan Chen and Chih-Cheng Hsueh

*Department of Information Engineering, I-Shou University,*

*Kaohsiung Country, Taiwan, 840, R.O.C*

**Abstract**

In this paper, we present a novel quantum secret key encryption algorithm based on quantum discrete logarithm problem (QDLP). QDLP is to find the exponent x from $C = \sum_i \left| g_i^x \bmod p \right\rangle$ if C, $\sum_i \left| g_i \right\rangle$ and the prime p are given, where $g_i = g^i \bmod p$ for generator g According to our knowledge, QDLP cannot be solved by Shor's quantum algorithm. Our algorithm assumes that the sender and the receiver share the secret key x. If the sender wants to send the message y to the receiver, the receiver first uses the secret key x to construct a quantum system $\sum_i \left| i \right\rangle \left| g_i^x \bmod p \right\rangle$. Then, the receiver sends the quantum channel $\sum_i \left| g_i^x \bmod p \right\rangle$ to the sender. After receiving $\sum_i \left| g_i^x \bmod p \right\rangle$, the sender uses the secret key x and the message y to generate ciphertext $\sum_i \left| x g_i^y \bmod p \right\rangle$. The receiver then construct a quantum system $\sum_i \left| i \right\rangle \left| x g_i^y \bmod p \right\rangle$. According to the secret key x, the receiver can compute and obtain final quantum system $\sum_i \left| i \right\rangle \left| g^y \bmod p \right\rangle$. Then, the receiver gets the value $g^y \bmod p$ after measuring. Finally, the receiver can successfully recover the message y by using

Shor's quantum algorithm. Obviously, the quantum system will be broken after transferring messages. But, the secret key x can still be used repeatedly in our algorithm.

## 1. Introduction

In quantum cryptography [1], secretly transferring messages between two entries is an important issue. Obviously, two entries can use quantum key distribution protocol [2] to obtain the secret key. With the secret key, two entries can use the one-time pads method in classical cryptography to secretly transfer the message. However, the secret key can be used once. If we want to use the secret key repeatedly, we must design a quantum secret key encryption algorithm. According to [4], Brassard and Bennett's key distribution protocol can be modified as the secret key encryption algorithm. Assume that the sender and the receiver share n secret bases. If the sender wants to transfer the message of n classical bits, he must generate n quantum states, viewed as ciphertexts. The i-th quantum state only depends on the i-th basis and the i-th classical bit of the message. After receiving the i-th quantum state, the receiver can use the i-th basis to measure the state and obtains the i-th classical bit of the message. Therefore, the modified Brassard and Bennett's protocol with shared bases can secretly transfer the message.

However, the secret key in the modified Brassard and Bennett's protocol is traceable due to the following attack. Assume that an attacker pretends the sender. He randomly selects a message of n classical bits and n bases. He generates n quantum states by using n bases and n classical bits, and sends them to the receiver. The receiver discovers n classical bits by measuring these quantum states with his n secret bases. Assume that the attacker obtains the discovered classical bits after transformation. If the i-th classical bit of the message is different from the i-th discovered classical bit, the attacker learns that the i-th secret basis is another one different from his i-th selected basis. Thus, the attacker can recover $\frac{n}{4}$ secret bases on average. To avoid this attack, this paper presents a novel quantum secret key encryption algorithm based on the quantum discrete logarithm problem.

In this paper, we first define the quantum discrete logarithm problem (QDLP). QDLP is similar to classical discrete logarithm problem (DLP). DLP is to find the exponent x from $C = g^x \mod p$ if C, g, and p are known. QDLP differs from DLP in that the base g is changed into a superposition $\sum_i |g_i\rangle$. Thus, QDLP is to find x from $C = \sum_i |g_i^x \mod p\rangle$ if C, $\sum_i |g_i\rangle$, and p are given. According to our knowledge, QDLP cannot be solved by Shor's algorithm [3]. Based on QDLP, we present a novel quantum secret key encryption algorithm. Assume that the receiver, Alice, and the sender, Bob, share the secret key x. Then, Bob wants to send a message y to Alice.

Thus, Alice prepares two quantum registers to construct her quantum channel. Let p and q be large primes satisfying $q | p-1$. She generates n-1 states $|i\rangle$ in first register, where i from 1 to q-1, and then uses secret key x to compute $\sum_{i=1}^{q-1}|g^{ix \bmod q} \bmod p\rangle$ which will be stored in second register. Thus, Alice has $\sum_{i=1}^{q-1}|i\rangle|g_i^{x \bmod q} \bmod p\rangle$, where $g_i = g^i \bmod p$. Alice sends her quantum channel $\sum_{i=1}^{q-1}|g_i^{x} \bmod p\rangle$ to Bob. After Bob gets Alice's quantum channel, Bob computes the ciphertext $\sum_{i=1}^{q-1}|xg_i^{xyx_q^{-1}} \bmod p\rangle = \sum_{i=1}^{q-1}|xg_i^{y} \bmod p\rangle$, where y is message and $x_q^{-1}$ satisfies $x_q^{-1} x \bmod q = 1$. Then, Bob sends it to Alice. After Alice receives Bob's ciphertext, she constructs $\sum_{i=1}^{q-1}|i\rangle|xg_i^{y} \bmod p\rangle$ by combining her original first register. She uses $x_p^{-1}$, the inverse of x modulo p, and the inverse of i to compute $\sum_{i=1}^{q-1}|i\rangle|x_p^{-1}xg_i^{i^{-1}y} \bmod p\rangle = \sum_{i=1}^{q-1}|i\rangle|g^y \bmod p\rangle$. Then, Alice obtains $g^y \bmod p$ by measuring the second register. At last, Alice obtains the message y by Shor's quantum algorithm [3]. According to the properties of quantum system, the quantum states will disappear after measurement. Thus, the quantum channel is used once for a message. Although the quantum channel is temporarily generated, the secret key x can be used repeatedly.

The remainder of this paper is organized as follows. In Section 2, we define the quantum discrete logarithm problem. Section 3 presents a novel quantum secret key encryption algorithm. We discuss and analyze our algorithm in Section 4. Finally, we draw the conclusions in Section 5.

**2. Quantum discrete logarithm problem**

In this section, we define the quantum discrete logarithm problem in Definition 1 and its general case in Definition 2.

***Definition 1***: (Quantum Discrete Logarithm Problem, QDLP)

Let p and q be two known large primes satisfying q|p-1. We can find g with order q modulo p. The quantum discrete logarithm problem is to find x from the superposition $\frac{1}{\sqrt{q-1}}\sum_{i=1}^{q-1}|i\rangle|g_i^{x \bmod q} \bmod p\rangle$, where $g_i = g^i \bmod p$.

The quantum discrete logarithm is similar to the classical discrete logarithm problem because $\sum_{i=1}^{q-1}|g_i \bmod p\rangle$ and p are known. But, the classical discrete logarithm problem is tractable due to Shor's algorithm. Shor's algorithm can discover x from $C = g^x \bmod p$. Shor prepares three quantum registers. He puts the first two registers in the uniform superposition of all $|a\rangle$ and $|b\rangle$ (mod p-1), and computes $g^a C^{-b} \bmod p$ in the third register. He has $\frac{1}{p-1}\sum_{a=0}^{p-2}\sum_{b=0}^{p-2}|a\rangle|b\rangle|g^a C^{-b} \bmod p\rangle$. Further, Shor can find the discrete logarithm x with two modular exponentiations and two quantum Fourier transforms. But, how to solve the quantum discrete logarithm problem by Shor's algorithm? According to Shor's algorithm, we add one quantum register that put the uniform superposition of all $|g_j\rangle$, where $j \in [1..p-2]$, and then

replace g with $\sum_{i=1}^{q-1}|g_{ji} \bmod p\rangle$, where $g_{ji} = g_j^i \bmod p$. Then we have

$$\frac{1}{(p-1)\sqrt{(q-1)(p-2)}} \sum_{i=1}^{q-1}\sum_{j=1}^{p-2}\sum_{a=0}^{p-2}\sum_{b=0}^{p-2}|a\rangle|b\rangle|g_j\rangle|g_{ji}^a C_{ji}^{-b} \bmod p\rangle,$$

where $C_{ji} = g_{ji}^x \bmod p$. Because the generator $g_{ji}$ is unknown, we must try all possible values $\sum_{j=1}^{p-2}|g_j\rangle$. Moreover, according to Shor's algorithm, we must also know the values of $C_{ji}$. Thus, it is difficult to obtain the discrete logarithm by Shor's algorithm.

In the following, we present the general case of QDLP in Definition 2 by modifying the first quantum register of Definition 1.

***Definition 2***: (General case of QDLP)

Given a large prime p, we can find g with order r modulo p. Let $a_0$, $a_1$, ..., $a_{k-1}$ are integers, less than r, satisfying GCD($a_i$, r)=1, where $i \in [0..k-1]$. The general case of QDLP is to find x from the superposition $\frac{1}{\sqrt{k}}\sum_{i=0}^{k-1}|a_i\rangle|g_{a_i}^x \bmod p\rangle$, where $g_{a_i} = g^{a_i} \bmod p$.

The general case of QDLP is more difficult than QDLP if $a_0$, $a_1$, ..., $a_{k-1}$ are unknown.

**3. Our algorithm**

In this section, we first introduce the parameters in our algorithm. We choose two large primes p and q satisfying q|p-1. Then, we find $g \in Z_p^*$ satisfying $g^q \bmod p = 1$.

Assume that the sender, Bob, wants to send the message y to the receiver, Alice. They must share the secret key $x \in Z_q^*$ whose inverse $x_q^{-1}$ satisfying $x_q^{-1} x \mod q = 1$ and inverse $x_p^{-1}$ satisfying $x_p^{-1} x \mod p = 1$. Bob and Alice perform the following steps to transfer the message y.

Step 1:

Alice prepares two quantum registers $\frac{1}{\sqrt{q-1}} \sum_{i=1}^{q-1} |i\rangle |0\rangle$. Alice uses the secret key x to compute the exponent of g to the power of ix and store it in second register, i.e. $\frac{1}{\sqrt{q-1}} \sum_{i=1}^{q-1} |i\rangle |g^{ix \mod q} \mod p\rangle$. Let $g_i = g^i \mod p$. This quantum system is rewritten as $\frac{1}{\sqrt{q-1}} \sum_{i=1}^{q-1} |i\rangle |g_i^{x \mod q} \mod p\rangle$. Let the quantum channel be $|\Phi\rangle_A = \sum_{i=1}^{q-1} |g_i^{x} \mod p\rangle$. Alice then sends $|\Phi\rangle_A$ to Bob.

Step 2:

After receiving $|\Phi\rangle_A$, Bob uses the message y, the secret key x and the inverse $x_q^{-1}$ to compute the ciphertext $|\Phi\rangle_C = \sum_{i=1}^{q-1} |xg_i^{xyx_q^{-1}} \mod p\rangle = \sum_{i=1}^{q-1} |xg_i^{y} \mod p\rangle$. Then, Bob sends $|\Phi\rangle_C$ to Alice.

Step 3:

After receiving $|\Phi\rangle_C$, Alice can obtain a quantum system $|\Phi\rangle_{D_1} = \frac{1}{\sqrt{q-1}} \sum_{i=1}^{q-1} |i\rangle |xg_i^{y} \mod p\rangle$ because she holds the original first quantum

register. Then, Alice uses the inverse $x_p^{-1}$ to compute

$$|\Phi\rangle_{D_2} = \frac{1}{\sqrt{q-1}}\sum_{i=1}^{q-1}|i\rangle\left|x_p^{-1}xg_i^y \bmod p\right\rangle = \frac{1}{\sqrt{q-1}}\sum_{i=1}^{q-1}|i\rangle\left|g_i^y \bmod p\right\rangle.$$

Finally, Alice computes the second register to power of the inverse $i^{-1}$ and obtains

$$|\Phi\rangle_{D_3} = \frac{1}{\sqrt{q-1}}\sum_{i=1}^{q-1}|i\rangle\left|g_i^{yi^{-1}} \bmod p\right\rangle = \frac{1}{\sqrt{q-1}}\sum_{i=1}^{q-1}|i\rangle\left|g^y \bmod p\right\rangle.$$

Because $i<q$ and q is a prime, each value i has the inverse $i^{-1}$ satisfying $i \times i^{-1} \bmod q = 1$. Thus, Alice can measure the second register and obtains the result $g^y \bmod p$. Because g and p are known, Alice easily discovers y by using Shor's algorithm.

*Example:*

Let p=11, q=5, g=3 and secret key x=3. Assume that Bob wants to send message y=3 to Alice. Then, they perform Step 1 to Step 3 as follows.

Step 1:

Alice generates the quantum channel

$$\frac{1}{\sqrt{4}}\sum_{i=1}^{4}|i\rangle\left|g_i^{x \bmod 5} \bmod 11\right\rangle$$

$$= \frac{1}{2}(|1\rangle\left|3^{1\times 3 \bmod 5} \bmod 11\right\rangle + |2\rangle\left|3^{2\times 3 \bmod 5} \bmod 11\right\rangle + |3\rangle\left|3^{3\times 3 \bmod 5} \bmod 11\right\rangle$$

$$+ |4\rangle\left|3^{4\times 3 \bmod 5} \bmod 11\right\rangle)$$

$$= \frac{1}{2}(|1\rangle|5\rangle + |2\rangle|3\rangle + |3\rangle|4\rangle + |4\rangle|9\rangle).$$

Alice sends $|\Phi\rangle_A = |5\rangle + |3\rangle + |4\rangle + |9\rangle$ to Bob.

Step 2:

Bob uses y=3, x=3 and $x_q^{-1} = 2$ to compute the ciphertext

$$|\Phi\rangle_c = |3 \times 5^{3 \times 2 \bmod 5} \bmod 11\rangle + |3 \times 3^{3 \times 2 \bmod 5} \bmod 11\rangle + |3 \times 4^{3 \times 2 \bmod 5} \bmod 11\rangle$$
$$+ |3 \times 9^{3 \times 2 \bmod 5} \bmod 11\rangle$$

$$= |4\rangle + |9\rangle + |1\rangle + |5\rangle.$$

Then, Bob sends $|\Phi\rangle_c$ to Alice.

Step 3:

Alice obtains the quantum system

$$|\Phi\rangle_{D_1} = \frac{1}{2}\sum_{i=1}^{4}|i\rangle|xg_i^y \bmod p\rangle = \frac{1}{2}(|1\rangle|4\rangle + |2\rangle|9\rangle + |3\rangle|1\rangle + |4\rangle|5\rangle).$$

Then, Alice uses $x_p^{-1} = 4$ to compute

$$|\Phi\rangle_{D_2} = \frac{1}{2}\sum_{i=1}^{q-1}|i\rangle|x_p^{-1}xg_i^y \bmod p\rangle$$

$$= \frac{1}{2}(|1\rangle|4 \times 4 \bmod 11\rangle + |2\rangle|4 \times 9 \bmod 11\rangle + |3\rangle|4 \times 1 \bmod 11\rangle + |4\rangle|4 \times 5 \bmod 11\rangle)$$

$$= \frac{1}{2}(|1\rangle|5\rangle + |2\rangle|3\rangle + |3\rangle|4\rangle + |4\rangle|9\rangle).$$

Finally, Alice computes the second register to power of the inverse of the first register and obtains

$$|\Phi\rangle_{D_3} = \frac{1}{2}\sum_{i=1}^{4}|i\rangle|g_i^{yi^{-1} \bmod q} \bmod p\rangle$$

$$= \frac{1}{2}(|1\rangle|5^1 \bmod 11\rangle + |2\rangle|3^3 \bmod 11\rangle + |3\rangle|4^2 \bmod 11\rangle + |4\rangle|9^4 \bmod 11\rangle)$$

$$= \frac{1}{2}(|1\rangle|5\rangle + |2\rangle|5\rangle + |3\rangle|5\rangle + |4\rangle|5\rangle).$$

Thus, Alice can measure the second register and obtains the output. Then, we have $g^y \bmod p = 5$. Because g=3 and p=11 is known, Alice easily computes y=3 by using Shor's algorithm.

## 4. Discussion and analysis

We divide four cases to discuss and analyze our algorithm in this section. Assume that an eavesdropper, called Nancy, exists in quantum channel.

Case 1:

Assume that an eavesdropper wants to get secret key x from quantum channel or ciphertext. He must face the QDLP. According to our knowledge, QDLP cannot be solved by Shor's quantum algorithm. Thus, our algorithm can prevent the attacks from quantum algorithms. Moreover, in our algorithm, the quantum channel is broke after measuring. However, Alice and Bob still securely share the secret key. Thus, we can use the secret key x more than once in our algorithm.

Case 2:

Assume that Nancy steals the ciphertext $|\Phi\rangle_c = \sum_{i=1}^{q-1} |xg_i^y \bmod p\rangle$. Nancy wants to discover the message y by using the measurement. Nancy can measures $|\Phi\rangle_c$ and obtains a value $C'$. Assume that $C' = xg_j^y \bmod p$, where $j \in \{1..q-1\}$. Because x and $g_j$ are unknown, Nancy cannot perform Shor's quantum algorithm to obtain the value y.

Case 3:

Assume that Nancy forges the Alice's quantum channel to get the message y. Nancy can construct the quantum system as $\frac{1}{\sqrt{q-1}}\sum_{i=1}^{q-1}|i\rangle|g_i^{x' \bmod q} \bmod p\rangle$, where $x'$ is the forged secret key. Nancy sends $|\Phi\rangle_{A'} = \frac{1}{\sqrt{q-1}}\sum_{i=1}^{q-1}|g_i^{x'} \bmod p\rangle$ to Bob. At last, Nancy gets the ciphertext $|\Phi\rangle_{c'} = \frac{1}{\sqrt{q-1}}\sum_{i=1}^{q-1}|xg_i^{x'yx_q^{-1}} \bmod p\rangle$. Then, Nancy uses the inverse $x_q'^{-1}$ to compute

$$|\Phi\rangle_{D_1'} = \frac{1}{\sqrt{q-1}}\sum_{i=1}^{q-1}|xg_i^{x_q'^{-1}x'yx_q^{-1}} \bmod p\rangle = \frac{1}{\sqrt{q-1}}\sum_{i=1}^{q-1}|xg_i^{yx_q^{-1}} \bmod p\rangle.$$

Finally, Nancy computes the second register to power of the inverse of the first register and obtains

$$|\Phi\rangle_{D_2'} = \frac{1}{\sqrt{q-1}}\sum_{i=1}^{q-1}|(xg_i^{yx_q^{-1}})^{i^{-1}} \bmod p\rangle = \frac{1}{\sqrt{q-1}}\sum_{i=1}^{q-1}|x^{i^{-1}}g^{yx_q^{-1}} \bmod p\rangle.$$

Because x is unknown, Nancy obtains the value $(x^{i^{-1}}g_i^{yx_q^{-1}})^{i^{-1}} \bmod p$. Thus, Nancy cannot decrypt the ciphertext to get message y by Shor's algorithm unless she knows the secret key x.

Case 4:

We can prepare the first quantum register as Definition 2. Alice prepares the quantum system $\frac{1}{\sqrt{k}}\sum_{i=0}^{k-1}|a_i\rangle|g_{a_i}^x \bmod p\rangle$. In the first quantum register, the value $a_i$ satisfies GCD($a_i$, r)=1, where r is the order of g modulo p. It is more difficult for Nancy to get the secret key x from quantum channel $\sum_{i=0}^{k-1}|g_{a_i}{}^x \bmod p\rangle$. Furthermore, it is also difficult to discover the message y from the ciphertext $|\Phi\rangle_c = \sum_{i=0}^{k-1}|xg_{a_i}{}^y \bmod p\rangle$. Thus, based on the general case of QDLP, the quantum secret key encryption algorithm is more secret.

## 5. Conclusion

In this paper, we define the quantum discrete logarithm problem (QDLP). QDLP differs from DLP in that the base g is changed into a superposition $\sum_i |g_i\rangle$. QDLP is to find x from $C = \sum_i |g_i{}^x \bmod p\rangle$ if C, $\sum_i |g_i\rangle$, and p are known. According to my knowledge, QDLP cannot be solved by Shor's algorithm. Based on QDLP, we present a novel quantum secret key encryption algorithm. Assume that Bob wants to send a message y to Alice. Alice prepares two quantum registers to construct her quantum channel. At last, Alice obtains the message y by Shor's quantum algorithm [3]. According to the properties of quantum system, the channel will be broken after measuring. Thus, the quantum channel is used once for a message. Although the quantum channel is temporarily constructed, the secret key x can be used repeatedly. Furthermore, we discuss that our quantum algorithm is based on the general case of QDLP. Because $\sum_i |g_{a_i} \bmod p\rangle$ is unknown, this algorithm is more secure.